\documentclass[11pt,a4paper]{article}

\usepackage{amsmath, amssymb, graphics}
\usepackage{graphicx}
\usepackage{epstopdf}
\usepackage{booktabs}

\usepackage{jheppub}

\input{epsf}

\numberwithin{equation}{section}

\title{Boundaries in the Moyal plane}
\author{H.~Falomir,}
\author{S.~A.~Franchino Vi\~{n}as,}
\author{P.~A.~G.~Pisani,}
\author{and F.~Vega}
\affiliation{IFLP-CONICET/ Departamento de F\'isica, Facultad de Ciencias Exactas\newline
Universidad Nacional de La Plata, C.C.\ 67 (1900), La Plata, Argentina\newline}
\emailAdd{falomir@fisica.unlp.edu.ar}
\emailAdd{safranchino@gmail.com}
\emailAdd{pisani@fisica.unlp.edu.ar}
\emailAdd{federicogaspar@gmail.com}

\abstract{
We study the oscillations of a scalar field on a noncommutative disc implementing the boundary as the limit case of an interaction with an appropriately chosen confining background. The space of quantum fluctuations of the field is finite dimensional and displays the rotational and parity symmetry of the disc. We perform a numerical evaluation of the (finite) Casimir energy and obtain similar results as for the fuzzy sphere and torus.
}


\notoc
\begin{document}

\maketitle

\section{Introduction}

The dynamics of quantum fields on noncommutative (NC) spaces \cite{Snyder:1946qz,Doplicher:1994zv} exhibits many fascinating properties owing to non-locality and to the existence of a minimal area \cite{Douglas:2001ba,Szabo:2001kg}. Among these properties is the possibility of describing field theories defined on compact NC spaces as matrix models and therefore representing them on finite dimensional Hilbert spaces. Quantum fields defined on the fuzzy sphere $S^2_F$ \cite{Madore:1991bw} and on the fuzzy torus $T^2_F$ \cite{Connes:1997cr} are well-known examples of these kind of models with a finite number of degrees of freedom.

The purpose of the present article is to study a free scalar field on a noncommutative generalization of the two dimensional disc. Apart from being compact, the disc has a boundary, whose definition in a noncommutative context is subtle. The NC plane is described by a pair of coordinates $(\hat{x}_1,\hat{x}_2)$ which satisfy the algebra $[\hat{x}_i,\hat{x}_j]=2i\theta\,\epsilon_{ij}$, where the noncommutativity (NC) parameter $\theta\in\mathbb{R}$ has dimensions of squared-length\footnote{Throughout this article we consider, without loss of generality, $\theta>0$.}. These commutation relations enforce the existence of a minimal area $4\pi\theta$ on the plane, which establishes that states highly localized in one direction must spread over large distances in the transverse one. In this context, where there are no simultaneous eigenstates of the operators $\hat{x}_1,\hat{x}_2$ and thus the usual notion of a point in the plane is lost, it is not evident how to generalize the concept of boundary.

In ordinary commutative field theory one can define quantum fields confined in a space-like subregion $\mathcal{R}$ by introducing in the Lagrangian on the whole space an interaction with an ``infinitely high step-function'', i.e.\ a confining background $V_{\mathcal{R}}$ which vanishes in $\mathcal{R}$ and becomes infinite outside $\mathcal{R}$. In this article we will generalize this procedure to the case where $\mathcal{R}$ is a disc in the Moyal plane. The Moyal plane provides a representation of the aforementioned commutation relations between coordinates as an algebra of functions in the ordinary commutative plane equipped with the Moyal product \cite{Groenewold:1946kp,Moyal:1949sk}:
\begin{equation}\label{mp}
  f(x)\star g(x):=f(x)\,e^{i\theta\,\epsilon_{ij}\overleftarrow{\partial}_i\overrightarrow{\partial}_j}\,g(x)\,.
\end{equation}
Note that $x_i\star x_j-x_j\star x_i=2i\theta\,\epsilon_{ij}$, as expected. This associative product can also be realized in terms of the Bopp's shifts
\begin{equation}\label{bs}
x_i\rightarrow x_i^{\pm}:=x_i\mp i\theta\,\epsilon_{ij}\partial_j
\end{equation}
as follows: $f(x)\star g(x)=f(x^{-})\,g(x)=g(x^{+})\,f(x)$.

We will therefore consider a free scalar field on the ordinary commutative plane coupled, by means of this noncommutative product, with a confining background $V_{\mathcal{R}}$, where $\mathcal{R}$ will be chosen as the disc of radius $R$. We will determine the spectrum of its one-loop oscillations and show that the quantum oscillations of the scalar field display the rotational and parity symmetry of the disc. At the same time, some peculiar properties show up.

Firstly, as in the case of $S^2_F$ and $T^2_F$, the space of quantum fluctuations is finite dimensional. In fact, from a semiclassical point of view, the constraint $\hat{x}^2_1+\hat{x}^2_2\leq R^2$ on the coordinates of the NC configuration space implies that the physical region of the system in phase space has finite volume. We will show that the scalar field on the NC disc is quantized on a Hilbert space whose dimension is given by the (finite) number of minimal cells contained in this volume.

On the other hand, we will show that the oscillation modes do not vanish abruptly at the boundary of the disc but decay exponentially beyond the radius $R$, which is to be expected for a boundary defined in terms of a non-local product. There exists another interesting behavior of the oscillation modes. As is well-known, in the commutative case the wave functions concentrate on the boundary of the disc as the angular momentum increases. On the contrary, we will see that at some value of the angular momentum the mean position of the quantum oscillations on the NC disc moves back towards the center of the disc.

\medskip

The study of vacuum oscillations on NC spaces has attracted interest in the context of Kaluza-Klein (KK) theories \cite{Gomis:2000sw}. In the usual commutative KK scenario the vacuum energy receives a contribution from the oscillations of the fields in the extra dimensions. In geometries where this contribution leads to attractive Casimir forces one assumes that at Planck scales --where nonperturbative quantum gravity is relevant-- new dynamics prevents the collapse of the compactified extra dimensions. Some examples have shown that the noncommutative character of KK extra dimensions could provide this stabilization mechanism. In fact, the oscillations of a self-interacting scalar field defined on $\mathbb{R}^{1+d}\times T^2_\theta\times\ldots\times T^2_\theta$ originate a Casimir force which, although attractive at tree level, could stabilize the radii of the compactified tori due to one-loop corrections \cite{Huang:2000qc}. By the same token, one-loop contributions to the Casimir energy for a scalar field defined on $\mathbb{R}^{1+d}\times T^D\times S^2_F$ can give a repulsive force that stabilizes the radius of the commutative torus\footnote{In this model the radius of $S^2_F$ is determined by the strength of a RR four-form background in a $D0$-brane scenario, in which the fuzzy sphere appears as a static solution \cite{Myers:1999ps}.} \cite{Huang:2002ic}.

It is also interesting to study the Casimir energy arising from KK oscillation modes as a possible source of dark energy. In fact, although a compactified $S^2$ gives a negative contribution to the energy density measured on our four-dimensional Minkoswki spacetime, this contribution can change sign if the extra dimensions are compactified on $S^2_F$ instead \cite{Fabi:2006rs}. Let us also mention that if the KK extra dimensions are compactified on $T^2_F$ the contribution to the vacuum energy is negative, so it cannot be considered as a source of dark energy \cite{Fabi:2007hx}.

With this motivation, in the last part of this article we study the Casimir energy of the scalar field on the NC disc. Since the Hilbert space of fluctuations is finite dimensional this energy is finite and only in the commutative limit $\theta\rightarrow 0$ the resulting expression diverges. Our results are similar to analogous calculations performed on $T^2_F$ \cite{Chaichian:2001pw} and on $S^2_F$ \cite{Demetrian:2002pm}.

\medskip

The first noncommutative generalization of the disc --the fuzzy disc $B^2_F$-- has already been defined by F.\ Lizzi, P.\ Vitale and A.\ Zampini \cite{Lizzi:2003ru,Lizzi:2005zx,Lizzi:2006bu} by means of a truncation of the expansion of operators in a creation/annihilation basis (see also \cite{Pinzul:2001qh,Balachandran:2003vm}). As a consequence, the oscillation modes span a finite-dimensional Hilbert space in which, following the Berezin formalism, the wave functions are defined as the expectation values of these truncated operators in coherent states. The NC algebra of functions on $B^2_F$ is therefore a finite subalgebra of the algebra of functions on the NC plane endowed with the so-called Voros product \cite{Voros:1989iu}. In this regard, our study of the NC disc could contribute to the comparison between the use of Moyal and Voros products in the formulation of field theories \cite{Hammou:2001cc,Galluccio:2008wk,Balachandran:2009kb,Balachandran:2009as,Varshovi:2012rh,Varshovi:2012yc} in the presence of boundaries.\footnote{A self-interacting field in terms of the Voros product has been studied numerically on the fuzzy disc \cite{Lizzi:2012xy}.} We also point out that whereas the definition of $B^2_F$ appears as the natural extension of the procedures used to define $S^2_F$ and $T^2_F$, our definition --which does not rely on the symmetries of the base manifold-- could in principle be applied to quantum fields confined in an arbitrary subregion of the NC plane.

The definition of boundaries in a NC plane by means of a confining background\footnote{The addition of a potential term to model boundaries in deformation quantization has been considered in \cite{Dias-Prata,Kryukov:2004ab}.} has already been used by C.\ D.\ Fosco et al.\ \cite{Fosco:2007tn} and by F.\ G.\ Scholtz et al.\ \cite{Scholtz:2007ig}. Let us point out some differences between the results in these articles and the ones discussed here. In \cite{Fosco:2007tn} the boundary is represented by a Moyal-interaction with a NC delta-function. Since the limit of an infinitely strong interaction is not explicitly worked out, a representation of the quantum fluctuations on a finite dimensional Hilbert space is not present in this analysis. In \cite{Scholtz:2007ig} a step-function is used as a confining background in the context of NC quantum mechanics. However, the wave functions are defined in terms of coherent states so that noncommutativity is represented, as in \cite{Lizzi:2003ru}, by the Voros product. Most importantly, since the Hamiltonian considered in \cite{Scholtz:2007ig} is not parity-invariant, the Hilbert space is not finite-dimensional.

The present article is organized as follows. In Section \ref{s2} we consider a scalar field on the Moyal plane interacting with a rotationally invariant background and present a basis of the Hilbert space which is convenient to describe the classical configurations of the field. In Section \ref{s3} we choose a particular background which, in some limit, defines our formulation of the NC disc. We find an explicit solution for the spectrum and the corresponding eigenfunctions and study some of their properties. In Section \ref{casim} we apply these results to study the Casimir energy on the NC disc. Finally, in Section \ref{s5} we draw our conclusions. The appendix \ref{thewell} contains a detailed calculation of the results which were obtained in Section \ref{s3} with a simpler but less rigorous argument.

\section{Scalar field in a rotationally invariant NC background}\label{s2}

Let us consider a real massive scalar field $\phi(t,x)$, where $t\in\mathbb{R}$ is Minkowski time and $x=(x_1,x_2)\in\mathbb{R}^2$ are coordinates on the Moyal plane, that interacts with a rotationally invariant background $V=V(r^2)$, where $r^2:=x_ix_i$. The corresponding action reads
\begin{equation}\label{action}
    S[\phi]=\frac12\int_{\mathbb{R}\times\mathbb{R}^2}dtdx\,\left\{(\partial_t \phi)^2-(\partial_x \phi)^2-m^2\,\phi^2-V\star\phi\star\phi\right\}\,.
\end{equation}
The $\star$-product is the associative Moyal product defined in eq.\ \eqref{mp}. Since under the integral sign the Moyal product of two functions can be replaced by the ordinary commutative product, the (quadratic) kinetic and mass terms are the same as in the commutative theory. For the same reason, the double Moyal product $V\star\phi\star\phi$ is invariant under cyclic permutations of the factors and there is no ordering ambiguity in the definition of the interaction term. As we will see below, this term gives rise to a non-local operator which acts symmetrically with respect to left- and right-Moyal multiplication and thus preserves parity invariance, which is the origin of the finite-dimensional representation of the quantum model.

The one-loop effective action of the model described by eq.\ \eqref{action} is given by
\begin{equation}
    \Gamma[\phi]=S[\phi]+\tfrac12 \hbar\, \log{{\rm Det}\left\{\delta^2S\right\}}\,,
\end{equation}
where $\delta^2S$ is the operator of one-loop quantum fluctuations, whose kernel is defined as the second functional derivative of $S[\phi]$ with respect to the field. Since the action is quadratic this operator is field independent and its zero modes determine the classical configurations of the scalar field. These oscillations modes constitute a basis of the Hilbert space of the operator-valued quantized field; we will see in Section \ref{s3} that for a scalar field in the NC disc this Hilbert space is finite-dimensional.

A straightforward calculation shows
\begin{equation}\label{olqf}
    \delta^2S:=-\partial_t^2-m^2-A\,,
\end{equation}
where we denote as
\begin{equation}\label{opa}
    A:=-\partial^2_x+\tfrac12\, V(x_i^+x_i^+)+\tfrac12\, V(x_i^-x_i^-)
\end{equation}
its spatial part, with $x^{\pm}_{i}$ the operators (or coordinates in phase space) defined in eq.\ \eqref{bs}. The last two terms in the operator $A$ represent the left- and right-Moyal multiplication by the background $V(r^2)$.

One-loop physical quantities related to the scalar field $\phi(t,x)$ can be derived from the spectrum of $A$. To diagonalize this non-local operator we will make use of its rotational invariance. Indeed, due to the rotational invariance of the background, it is convenient to define the non-Hermitian dimensionless operators
\begin{equation}
    a_{\pm}:=\frac{1}{2\sqrt{\theta}}(x^\pm_1 \mp i x^\pm_2)\,,
\end{equation}
which satisfy the algebra of a pair of uncoupled creation/annihilation operators whose non-vanishing commutators are
\begin{equation}
    [a_+,a^\dagger_+]=[a_-,a^\dagger_-]=1\,.
\end{equation}
In terms of the corresponding number operators $N_{\pm}:=a^\dagger_\pm a_\pm$ the angular momentum reads
\begin{equation}\label{L}
    L:=-i\epsilon_{ij}x_i\partial_j=N_+-N_-\,.
\end{equation}
Note that $L$ is also the generator of rotations in the shifted coordinates, that is, $[L,x^\pm_i]=i\epsilon_{ij}\,x^{\pm}_{j}$. As eq.\ \eqref{L} shows, $a^\dagger_\pm$ and $a_\pm$ create and annihilate excitations with circular polarization.

Hereafter we take $l\in\mathbb{Z}^+$ as the absolute value of the angular momentum and we denote by $\mathcal{F}^{\pm l}$ the subspaces of states with definite angular momenta $\pm l$. The Hilbert space of square integrable functions on the plane can therefore be written as $L_2(\mathbb{R}^2)=\mathcal{F}^{0}\oplus_{l=1}^\infty\mathcal{F}^{l}\oplus_{l=1}^\infty\mathcal{F}^{-l}$. The orthonormal Fock basis of $\mathcal{F}^{\pm l}$ is given by the functions
\begin{equation}\label{basis}
    \phi^{\pm l}_{n}(x)=\frac{1}{\sqrt{(n+l)!\,n!}}\,(a^\dagger_{\pm})^{n+l}(a^\dagger_{\mp})^{n}\phi^0_0(x)\,,
\end{equation}
respectively, where $\phi^0_0(x)$ is defined by $a_+\,\phi^0_0(x)=a_-\,\phi^0_0(x)=0$. Note that these vectors are eigenfunctions of the operators $N_{\pm}$ with eigenvalues $n+ l$ and $n$. From eq.\ \eqref{basis} one straightforwardly obtains \cite{G-R}
\begin{eqnarray}\label{basisex}
    \phi^{\pm l}_n(x)=\frac{(-1)^n}{\sqrt{\pi\theta^{l+1}}}\frac{\sqrt{n!}}{\sqrt{(n+l)!}}
    \ r^{l}e^{-\frac{r^2}{2\theta}}L^{l}_n(r^2/\theta)\ e^{\pm il\varphi}\,,
\end{eqnarray}
where $L^{l}_n(\cdot)$ are the generalized Laguerre polynomials and $r,\varphi$ are the polar coordinates of $x\in\mathbb{R}^2$.

In terms of the creation/annihilation operators, the operator $A$ reads
\begin{align}\label{opan}
    \theta\, A&=N_++N_-+1-a^\dagger_+a^\dagger_--a_+a_-+\\\nonumber
    &\mbox{}+\frac\theta2\left\{ V(2\theta(2N_++1))+V(2\theta(2N_-+1))\right\}\,.
\end{align}

Let us make some remarks regarding parity transformation in this Hilbert space. In two space dimensions this transformation is implemented by changing the sign of one coordinate or, upon a rotation in $\pi/2$, by interchanging both coordinates. This amounts to the interchange $x^{\pm}_1\leftrightarrow x^{\mp}_2$ (see eq.\ \eqref{bs}), which corresponds to $a_{\pm}\leftrightarrow \mp ia_\mp$ and, consequently, to the interchange $N_+\leftrightarrow N_-$. Expression \eqref{opan} then shows that $A$ is invariant under parity transformations.

Since $[A,L]=0$ (see eqs.\ \eqref{L} and \eqref{opan}) we can diagonalize $A$ on each subspace $\mathcal{F}^{\pm l}$ of definite angular momentum. Due to parity-invariance, its spectra on $\mathcal{F}^l$ and $\mathcal{F}^{-l}$ coincide and the corresponding eigenfunctions are related by complex conjugation. This two-fold degeneracy of the spectrum is consistent with the invariance under parity and rotations of the original action $S[\phi]$.

\section{The NC disc}\label{nc-well}\label{s3}

Next, we apply the preceding results to confine a scalar field in a NC disc. To that end, we consider in the action \eqref{action} the $\Lambda\rightarrow\infty$ limit of the background
\begin{equation}\label{bg}
    V(r^2):=\frac{2\Lambda}{\theta}\,\Theta(r^2-R^2)\,,
\end{equation}
where $\Theta(\cdot)$ is the step-function defined as $1$ if its argument is non-negative. This background represents a circular well of radius $R$ and height $2\Lambda$ (in units of $\theta^{-1}$).

Replacing \eqref{bg} into eq.\ \eqref{opan} we obtain the following expression for the spatial part $A^\Lambda_N$ of the operator of quantum fluctuations:
\begin{equation}\label{opan2}
    \theta\, A^\Lambda_N=N_++N_-+1-a^\dagger_+a^\dagger_--a_+a_-
    +\Lambda\,\left\{ \Theta(N_+-N)+\Theta(N_--N)\right\}\,,
\end{equation}
where $N$ is the ceiling function
\begin{equation}
N:=\lceil R^2/4\theta-1/2 \rceil\,,
\end{equation}
which represents the lowest integer greater or equal than $R^2/4\theta-1/2$. This positive integer $N$ is, roughly, the integer part of the quotient between the area of the well $\pi R^2$ and the fundamental area $4\pi\theta$. In consequence, the integer $N$ measures the noncommutativity of the model, being large for the ``almost commutative'' case $\theta\ll R^2$. Note that, as opposed to the commutative case, the step-functions on the {\small R.H.S.} of eq.\ \eqref{opan2} do not depend uniquely on the value of the radial coordinate but represent instead constraints on the Fock space. Our choice of a basis of eigenstates of the operators $N_{\pm}$, given by expression \eqref{basis}, allows a direct implementation of these constraints.

As already mentioned, we can determine the spectrum of $A^\Lambda_N$ on each subspace $\mathcal{F}^{\pm l}$ of angular momentum $\pm l$, separately. Consequently, the solutions $\psi^{\pm l}_\lambda(x)\in\mathcal{F}^{\pm l}$ to the eigenvalue equation
\begin{equation}\label{eigequ}
    \left(A^\Lambda_N-\frac{\lambda}{\theta}\right)\psi^{\pm l}_\lambda(x)=0
\end{equation}
admit the following expansion (see eq.\ \eqref{basisex}):
\begin{equation}\label{expansion}
    \psi^{\pm l}_\lambda(x)=\sum_{n=0}^\infty c^{l}_n(\lambda)\ \phi^{\pm l}_n(x)\,,
\end{equation}
where, due to parity invariance, the coefficients $c^l_n(\lambda)$ do not depend on the sign of the angular momentum.

We define the spatial part $A^\infty_N$ of the operator of quantum fluctuations on the NC disc as the limit $\Lambda\rightarrow\infty$ of $A^\Lambda_N$. Following the lines of \cite{Scholtz:2007ig}, one can determine the spectrum of $A^\infty_N$ by replacing expansion \eqref{expansion} into eq.\ \eqref{eigequ} and then considering the limit $\Lambda\rightarrow\infty$ of the resulting eigenvalues and eigenfunctions. The result of this rather laborious --though straightforward-- procedure is shown in Appendix \ref{thewell}. Nevertheless, a simpler inspection of the consequences of the $\Lambda\rightarrow\infty$ limit in eq.\ \eqref{eigequ} will also lead us to the eigenvalues and eigenfunctions of $A^\infty_N$.

First of all, it can already be seen from the last two terms in expression \eqref{opan2} that in the limit $\Lambda\rightarrow\infty$ the components of the solution $\psi^{\pm l}_\lambda(x)$ in the direction of the basis vectors corresponding to $N_{\pm}\geq N$ must tend to zero; otherwise, the step functions would give an infinite contribution to eq.\ \eqref{eigequ}. As a consequence, the relevant Hilbert space only contains states with $N_{\pm}=0,1,2,\ldots,N-1$ and has therefore dimension $N^2$. This is the first remarkable feature of the NC disc: the quantum oscillations of the scalar field take values on a finite-dimensional Hilbert space, in correspondence with similar well-known results obtained for other fuzzy compact spaces. Note the relevance of parity-invariance in this result; this symmetry implies the presence of both operators $N_+$ and $N_-$ in expression \eqref{opan2} and, consequently, leads to the constraints $0\leq N_+<N$ and $0\leq N_-<N$, which determine the dimension of the Hilbert space.

Recall now that the basis vectors $\phi^{\pm l}_n(x)$ (defined in eq.\ \eqref{basis}) are eigenfunctions of the operators $N_{\pm}$ with eigenvalues $n+ l$ and $n$. Let us translate condition $N_{\pm}<N$ to the corresponding condition for the integers $l$ and $n$. On the one hand, it implies that eq.\ \eqref{eigequ} has solutions only for $l<N$, i.e.\ the absolute value $l$ of the angular momentum of the quantum oscillations of the scalar field on the NC disc is bounded by $N$. On the other hand, it also implies that for fixed $l$ the coefficients $c^l_n(\lambda)$ in expansion \eqref{expansion} tend to zero as $\Lambda\rightarrow\infty$ for $n\geq N-l$. As a consequence, in each space $\mathcal{F}^{\pm l}$ (with $l<N$) the subspace of solutions of eq.\ \eqref{eigequ} in the limit $\Lambda\rightarrow\infty$ is ($N-l$)-dimensional. On the whole, there are $(N-l)$ solutions in each space of angular momentum $\pm l$, with $l<N$; this gives $N^2$ solutions, as already mentioned.

To determine the coefficients $c^l_ n(\lambda)$, we replace expansion \eqref{expansion} into eq.\ \eqref{eigequ}. As already noted, for $l<N$ and $n<N-l$ the step functions in eq.\ \eqref{opan2} do not contribute, so for these values of $l$ and $n$ we get
\begin{align}\label{recc}
    (l+1-\lambda)\,c^l_0(\lambda)-\sqrt{l+1}\,c^l_1(\lambda)&=0\,,\\\nonumber
    (2n+l+1-\lambda)\,c^l_n(\lambda)-\sqrt{(n+l)n}\,c^l_{n-1}(\lambda)-\sqrt{(n+l+1)(n+1)}\,c^l_{n+1}(\lambda)&=0\,.
\end{align}
If we compare these recurrence relations with the following functional relations between the associated Laguerre polynomials \cite{G-R}
\begin{align}\label{recLag}
    (l+1-\lambda)\,L^l_0(\lambda)-L^l_1(\lambda)&=0\,,\\\nonumber
    (2n+l+1-\lambda)\,L^l_n(\lambda)-(n+l)\,L^l_{n-1}(\lambda)-(n+1)\,L^l_{n+1}(\lambda)&=0\,,
\end{align}
we obtain the following solution for the coefficients $c^l_n(\lambda)$:
\begin{equation}\label{cl}
    c^l_n(\lambda)=\frac{\sqrt{n!}}{\sqrt{(n+l)!}}\,L^l_n(\lambda)\qquad {\rm for\ }l<N {\rm\ and\ }n<N-l\,.
\end{equation}
Since these solutions do not depend on $\Lambda$ they hold true also in the limit $\Lambda\rightarrow\infty$. Yet recurrence relations \eqref{recc} and \eqref{recLag} for $n=N-l-1$ imply that the coefficient $c^l_{N-l}(\lambda)$ is also given by eq.\ \eqref{cl}. However, as argued above, this coefficient must vanish in the limit $\Lambda\rightarrow\infty$. In terms of the corresponding Laguerre polynomial this condition reads
\begin{equation}\label{spec}
    L^l_{N-l}(\lambda)=0
\end{equation}
and determines the spectrum of the operator $A^\infty_N$.

\medskip

In sum, the operator $A^\infty_N$ --corresponding to the spatial part of the quantum fluctuations of a scalar field on a NC disc (see eq.\ \eqref{opa})-- is defined on an $N^2$-dimensional subspace of $\mathcal{F}^0\oplus_{l=1}^{N-1}\mathcal{F}^l\oplus_{l=1}^{N-1}\mathcal{F}^{-l}$. In the subspaces $\mathcal{F}^{\pm l}$ the eigenvalues are given (in units of $\theta^{-1}$) by the $(N-l)$ zeros $\lambda^l_k$ of the Laguerre polynomial $L^l_{N-l}(\lambda)$:
\begin{equation}\label{ncspec}
    \frac{1}{\theta}\,\lambda^l_k\qquad {\rm with\ }l=0,1,\ldots,N-1\ {\rm and\ }k=1,2,\ldots,N-l\,.
\end{equation}
The corresponding orthonormal eigenfunctions $\psi^{\pm l}_k(x)$ are obtained by replacing expression \eqref{cl} into expansion \eqref{expansion}:
\begin{equation}\label{eigenfunctions}
    \psi^{\pm l}_k(x)=\mathcal{N}^l_k\ e^{\pm il\varphi}\ r^{l}e^{-\frac{r^2}{2\theta}}
    \ \sum_{n=0}^{N-l-1}(-1)^n\frac{n!}{(n+l)!}\, L^l_n(\lambda^l_k)L^{l}_n(r^2/\theta)\,,
\end{equation}
where the normalization constant is
\begin{equation}
    \mathcal{N}^l_k:=\frac{\sqrt{\lambda^l_k}}{\sqrt{\pi\theta^{l+1}}}\frac{\sqrt{(N-1)!}}{\sqrt{N(N-l)!}}
    \,\frac{1}{L^l_{N-l-1}(\lambda^l_k)}\,.
\end{equation}

\medskip

As remarked in the Introduction, the finite dimension $N^2$ of the Hilbert space can be understood from a semiclassical point of view in terms of the number of minimal cells contained in the physical region of phase space. Indeed, the constraint $\hat{x}^2_1+\hat{x}^2_2\leq R^2$ implies that the coordinates $x^\pm_i$ of classical phase space are restricted to a volume $(\pi\,R^2)^2$, whereas the commutation relations $[x^\pm_i,x^\pm_j]=\mp 2i\theta\,\epsilon_{ij}$ define minimal cells of volume $(4\pi\theta)^2$. Consequently, the number of cells contained in the physical volume is $(R^2/4\theta)^2\sim N^2$ in the semiclassical limit $\theta\ll R^2$.

\begin{figure}[t]
\centering
\begin{minipage}{.46\textwidth}
\centering
\includegraphics[height=40mm]{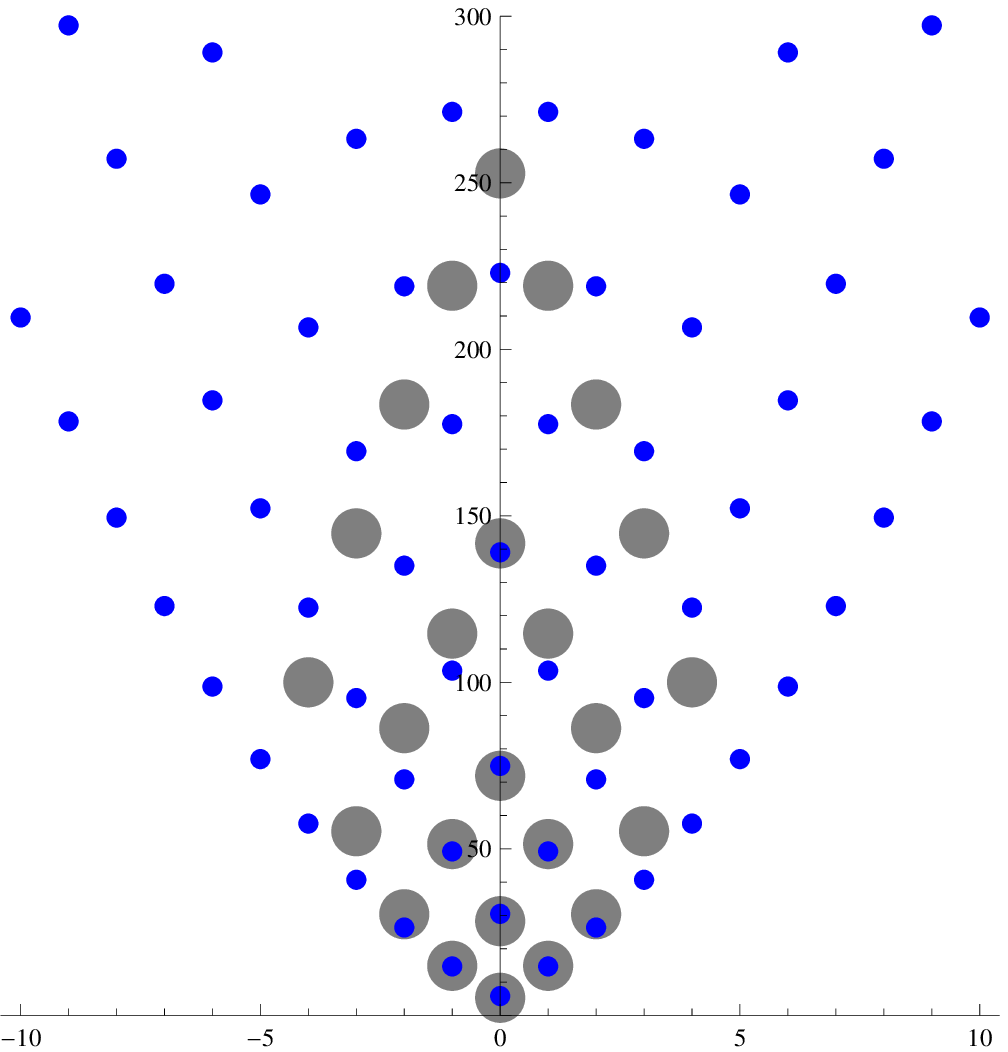}
\caption{Big dots represent the spectrum of $A^\infty_N$ for $\theta=0.05\,R^2$ ($N=5$) in terms of the angular momentum. Small dots correspond to the spectrum of the Laplacian on the commutative disc. $R$ is taken as the unit length.}
\label{n4}
\end{minipage}
\hspace{0.05\textwidth}
\begin{minipage}{.46\textwidth}
\centering
\includegraphics[height=40mm]{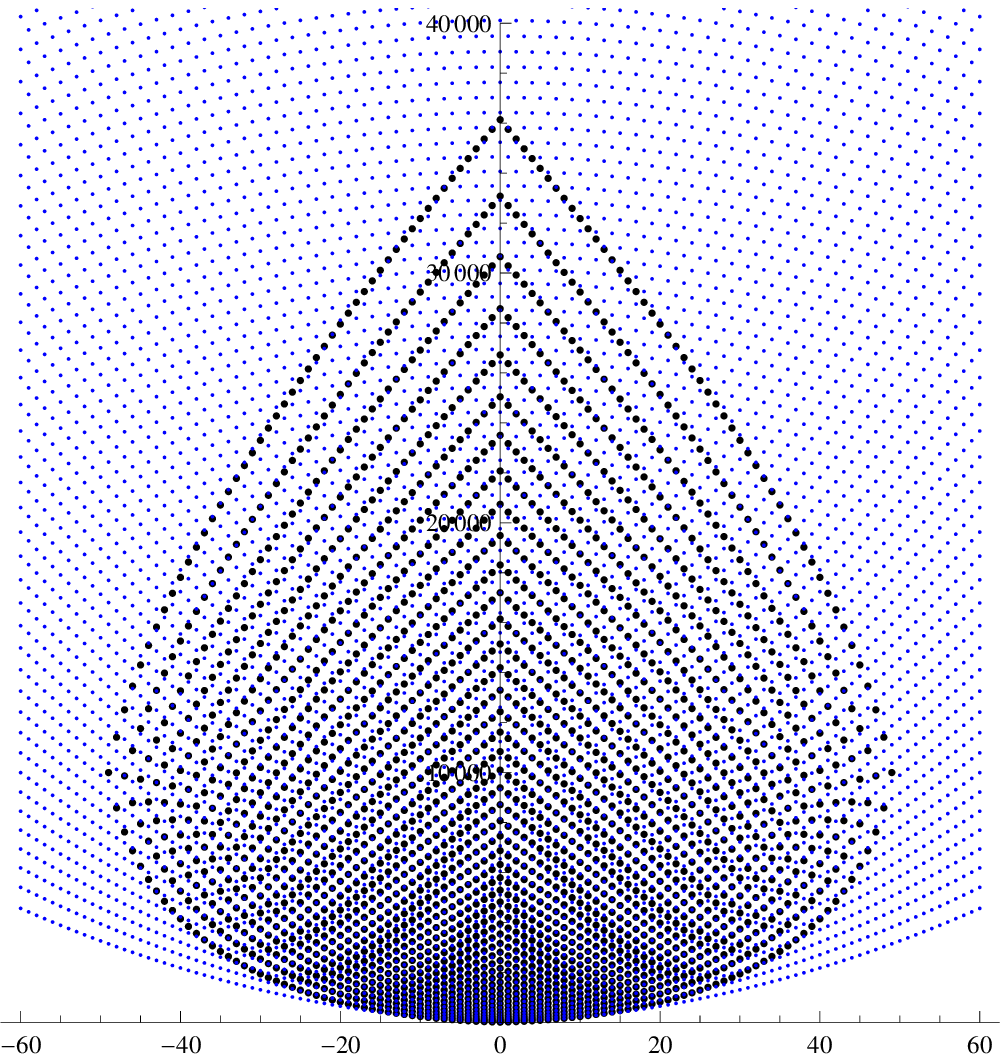}
\caption{Dots composing the ``droplet'' represent the spectrum of $A^\infty_N$ for $\theta=0.005\,R^2$ ($N=50$) in terms of the angular momentum. Dots of the background correspond to the spectrum of the Laplacian on the commutative disc. $R$ is taken as the unit length.}
\label{n49}
\end{minipage}
\end{figure}

\begin{table}[t]
\tiny
\centering
{\setlength{\doublerulesep}{1pt}
\begin{tabular}{@{}lrrrrrrrrr@{}}
\toprule\toprule
       & $N=1$    & $N=2$   & $N=3$   &  $N=4$   & $N=25$   & $N=50$   & $N=500$  & $N=1000$  & $N=\infty$\\[0.5ex]
       &$\theta=\frac{1}{6}$ &$\theta=\frac{1}{10}$&$\theta=\frac{1}{14}$&$\theta=\frac{1}{18}$ &$\theta=\frac{1}{100}$ &$\theta=\frac{1}{200}$ &$\theta=\frac{1}{2000}$ &$\theta=\frac{1}{4000}$ &$\theta=0$\\[0.5ex]
\midrule
\addlinespace[3pt]
$l=0$  &  6       & 5.85786 & 5.82084 &  5.80586 & 5.67048  & 5.7261   & 5.77741  & 5.7803   & 5.78319   \\
       &          & 34.1421 & 32.1199 &  31.4237 & 29.9011  & 30.1766  & 30.4409  & 30.4561  & 30.4713   \\
       &          &         & 88.0592 &  81.6592 & 73.591   & 74.1898  & 74.8126  & 74.8497  & 74.887    \\
       &          &         &         &  169.111 & 136.918  & 137.818  & 138.903  & 138.971  & 139.04    \\
       &          &         &         &          & $\cdots$ & $\cdots$ & $\cdots$ & $\cdots$ & $\cdots$    \\\hline
\addlinespace[2pt]
$l=1$  &          & 20      & 17.7513 &  16.8448 & 14.6892  & 14.6838  & 14.682   & 14.682   & 14.682    \\
       &          &         & 66.2487 &  59.4973 & 49.2995  & 49.2387  & 49.2187  & 49.2185  & 49.2185   \\
       &          &         &         &  139.658 & 103.859  & 103.589  & 103.5    & 103.5    & 103.499   \\
       &          &         &         &          & $\cdots$ & $\cdots$ & $\cdots$ & $\cdots$ & $\cdots$    \\\hline
\addlinespace[2pt]
$l=2$  &          &         & 42      &  36      & 26.9432  & 26.6484  & 26.4011  & 26.3878  & 26.3746   \\
       &          &         &         &  108     & 72.4898  & 71.6125  & 70.9214  & 70.8856  & 70.85     \\
       &          &         &         &          & $\cdots$ & $\cdots$ & $\cdots$ & $\cdots$ & $\cdots$    \\\hline
\addlinespace[2pt]
$l=3$  &          &         &         &  72      & 42.4899  & 41.5577  & 40.7882  & 40.7473  & 40.7065   \\
       &          &         &         &          & $\cdots$ & $\cdots$ & $\cdots$ & $\cdots$ & $\cdots$    \\\hline
\addlinespace[2pt]
$l=24$ &          &         &         &          & 2500     & 1181.11  & 903.655  & 893.022  & 882.714    \\\hline
\addlinespace[2pt]
$l=49$ &          &         &         &          &          & 10000    & 3305.12  & 3222.05  & 3144.17   \\\hline
\addlinespace[2pt]
$l=499$&          &         &         &          &          &          & $10^6$   & 362081   & 264041    \\\hline
\addlinespace[2pt]
$l=999$&          &         &         &          &          &          &          & $4\cdot10^6$  & $1.03562\cdot10^6$ \\
\bottomrule\bottomrule
\addlinespace[2mm]
\end{tabular}}
\caption{Spectrum of $A^\infty_N$ for different values of $\theta$ (in units of $R^2$) according to the absolute value $l$ of the angular momentum. For fixed $l$ the eigenvalues tend, as $N$ grows, to the eigenvalues of commutative case (shown in the last column). Note that $A^\infty_N$ admits $N^2$ eigenvalues.}
\label{NCvsC}
\end{table}

Figures \ref{n4} and \ref{n49} represent the spectrum of $A^\infty_N$ for two different values of the NC parameter, $\theta=0.05\,R^2$ and $\theta=0.005\,R^2$, corresponding to $N=5$ and $N=50$, respectively. The number of eigenvalues are $N^2=25$ and $N^2=2500$, respectively. The horizontal axes indicate the angular momentum, which --as shown in the figures-- run from $-N+1$ to $N-1$. In both cases the eigenvalues corresponding to $\theta=0$ are also displayed.

Let us mention that the spectrum given by expression \eqref{ncspec} coincides with the eigenvalues of the Laplacian in the fuzzy disc if one truncates the symbol of the field operator at a different integer as the one chosen in \cite{Lizzi:2003ru}. Of course, in our model this kind of arbitrariness also appears if one defines $\Theta(0)=0$ in eq.\ \eqref{bg}. Therefore, this ambiguity --which corresponds to changes of the order of $\theta$ in the area of the disc-- would be resolved from the specific microscopic behavior of the confining background in the vicinity of the boundary. As expected, it can be seen that the oscillation modes given by eq.\ \eqref{eigenfunctions} do not coincide with the eigenfunctions of the fuzzy Laplacian described in \cite{Lizzi:2005zx} since the latter are defined differently, in terms of expectation values in coherent states.

Next, we consider the commutative limit $\theta\rightarrow 0$ of the spectrum of the NC disc. Since the zeros of the Laguerre polynomials, $L^l_{N-l}(\lambda^l_k)=0$, and the zeros of the Bessel functions, $J_{l}(j^l_k)=0$, are related by \cite{A-S}
\begin{equation}\label{lagbes}
        \lambda^l_k=\frac{\left(j^l_k\right)^2}{4N-2l+2}+
        \frac{\left(j^l_k\right)^4+2(l^2-1)\left(j^l_k\right)^2}{3(4N-2l+2)^3}+
        O(N^{-5})\,,
\end{equation}
for fixed $l,k$ and $N\rightarrow\infty$, the eigenvalues of $A^\infty_N$ tend to $(j^l_k)^2/R^2$ (the eigenvalues of the commutative disc), for $N\gg l$. This is shown in Table \ref{NCvsC}, where one can see that as $N$ increases the eigenvalues of $A^\infty_N$ tend to the eigenvalues of the commutative disc as long as $l$ is not too large. Eq.\ \eqref{lagbes} shows that the eigenvalues of the commutative disc at fixed angular momentum can be approximated by the eigenvalues of the NC case for sufficiently large $N$. In fact, the lowest eigenvalue $\lambda^l_1$ for a fixed angular momentum $\pm l$ satisfies the bound $\lambda^l_1>(l+2)^2/(4N+2)$ \cite{Gatteschi,IS}. Therefore, all eigenvalues of the operator $A^\infty_N$ corresponding to this angular momentum are bounded by
\begin{equation}\label{cota}
    \frac{(l+2)^2}{R^2+2\theta}<\frac{\lambda^l_k}{\theta}\,.
\end{equation}
This inequality imposes an upper bound on the angular momentum that is probed according to the energy scale that is experimentally reached. Indeed, expression \eqref{cota} indicates that any measurement involving energies much smaller than a cutoff $E_{{\rm Planck}}$ only tests modes whose angular momenta satisfy $l\ll R\,\sqrt{E_{{\rm Planck}}}$. Being $l$ bounded by the experimental setup, eq.\ \eqref{lagbes} then shows that the spectrum of the NC disc uniformly approaches the commutative spectrum as $N\rightarrow \infty$. Subleading terms in eq.\ \eqref{lagbes} determine the experimental bounds on the NC parameter $\theta$.

\begin{figure}[t]
\centering
\begin{minipage}{.46\textwidth}
\centering
\includegraphics[height=45mm]{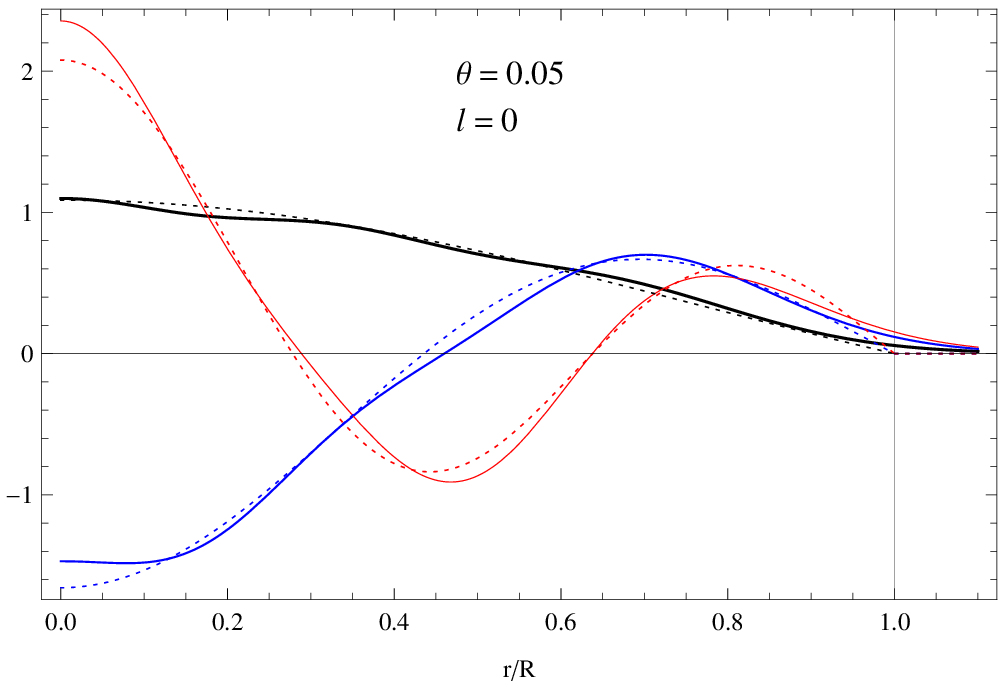}
\caption{Oscillation modes corresponding to the three lowest states ($k=1,2,3$) for vanishing angular momentum and $\theta=0.05$ (N=5) compared to their commutative counterparts (dashed curves). $R$ is taken as the unit length.}
\label{fig3}
\end{minipage}
\hspace{0.05\textwidth}
\begin{minipage}{.46\textwidth}
\centering
\includegraphics[height=50mm]{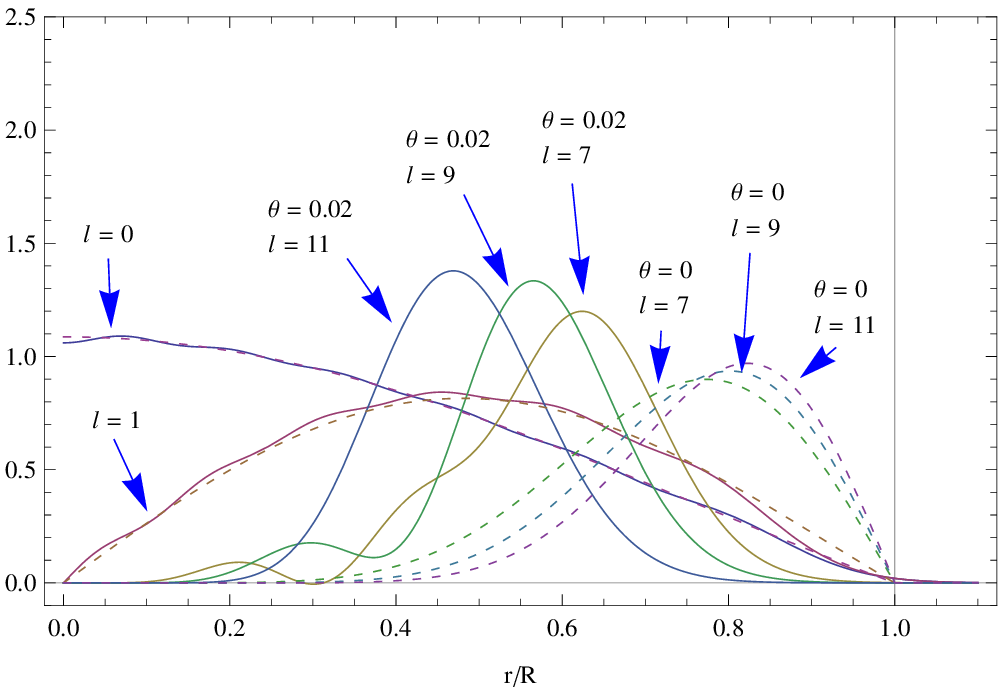}
\caption{Oscillation modes corresponding to the lowest states ($k=1$) for angular momentum $l=0,1,7,9,11$ and $\theta=0.02$ ($N=12$). Dashed curves represent their commutative counterparts. $R$ is taken as the unit length.}
\label{fig4}
\end{minipage}
\end{figure}

In Figure \ref{fig3} we plot the eigenfunctions given by eq.\ \eqref{eigenfunctions} corresponding to the three lowest states ($k=1,2,3$) with vanishing angular momentum, for $\theta=0.05\,R^2$ ($N=5$), together with the corresponding commutative modes. Contrary to the commutative case, in which the modes vanish abruptly at the radius of the disc $r=R$, the eigenfunctions of the NC case, though exponentially suppressed, penetrate the region $r>R$. One can also see that the approximation to the commutative case is worse as the number of nodes increases.

The oscillations of a scalar field confined in a NC disc manifest a peculiar phenomenon. As is well-known, the oscillation modes on the ordinary commutative disc concentrate closer to the boundary as the angular momentum increases. On the contrary, the wave functions of the NC case move towards the boundary as the angular momentum increases only for the first angular momenta but, at some point, as the angular momentum takes higher values the wave functions move back towards the center of the disc. Note that as $l\rightarrow N-1$ less terms in the sum in the {\small R.H.S.} of eq.\ \eqref{eigenfunctions} contribute to the wave function, reducing to just one term when $l=N-1$. Figure \ref{fig4} shows the lowest states (i.e.\ $k=1$) with angular momenta  $l=0,1,7,9,11$, for $\theta=0.02$ ($N=12$). Note that as $l$ takes the values $0,1,7$ the eigenfunctions move towards the boundary but as $l$ takes the values $7,9,11$ the eigenfunctions move away from the boundary. As we know, $l=11$ is the highest angular momentum for $N=12$. Dashed curves in Figure \ref{fig4} represent the corresponding eigenfunctions for the commutative disc which, as mentioned, approach the boundary monotonically as the angular momentum increases.

\begin{figure}[t]
\centering
\begin{minipage}{.46\textwidth}
\centering
\includegraphics[height=45mm]{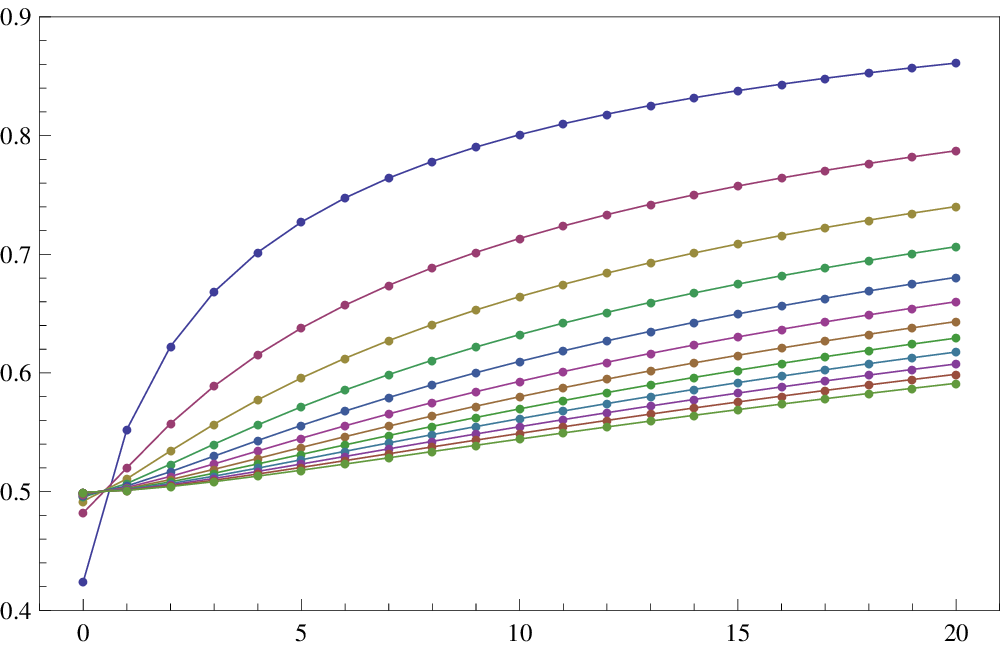}
\caption{Mean radius of the oscillation modes of the commutative case as a function of the angular momentum. Each line represents a different value of $k$, ranging (from top to bottom) from $k=1$ to $k=12$. $R$ is taken as the unit length.}
\label{fig5}
\end{minipage}
\hspace{0.05\textwidth}
\begin{minipage}{.46\textwidth}
\centering
\includegraphics[height=45mm]{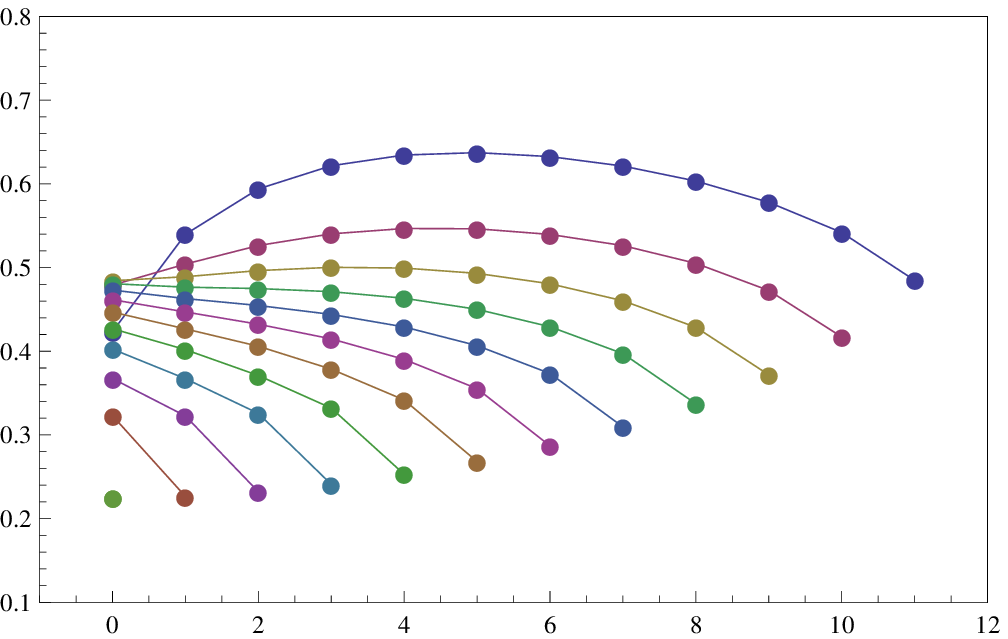}
\caption{Mean radius of the oscillation modes of the NC case ($N=12$ and $\theta=0.02$) as a function of the angular momentum. Each line represents a different value of $k$, ranging (from top to bottom) from $k=1$ to $k=12$. $R$ is taken as the unit length.}
\label{fig6}
\end{minipage}
\end{figure}

In order to further illustrate this phenomenon we consider the mean radius $\langle r\rangle$ of the oscillation modes as a function of the angular momentum. In Figure \ref{fig5} we plot $\langle r\rangle$ in the commutative case, for $0\leq l\leq 20$ and $k=1,2,\ldots,12$; each curve joins points corresponding to a fixed value of $k$ (i.e., with the same number of nodes). As expected, for fixed $k$, the mean radius grows monotonically as the angular momentum increases. In Figure \ref{fig6} we plot $\langle r\rangle$ in the NC disc for $\theta=0.02$ ($N=12$) so the absolute value of the angular momentum is restricted to $0\leq l\leq 11$. For each value of $l$, the oscillation modes are indexed by $k=1,2,\ldots,12-l$. Contrary to the commutative case, the figure shows that, for $k$ fixed and $l$ large enough, the mean radius of the oscillation mode decreases as the angular momentum increases.

Further insight into this result is provided by the small-$\theta$ expansion of the Moyal product (see eq.\ \eqref{mp})
\begin{equation}\label{pmp}
  f(x)\star g(x)\simeq f(x)g(x)+i\theta\,\epsilon_{ij}\,\partial_i f(x)\partial_j g(x)-
  \tfrac{\theta^2}{2}\,\epsilon_{ij}\epsilon_{kl}\,\partial^2_{ik}f(x)\partial^2_{jl}g(x)+\ldots
\end{equation}
Inserting this expansion into expression \eqref{opa} for the operator $A$ (acting on functions with angular momentum $\pm l$) we get to leading order in $\theta$
\begin{equation}\label{popal}
        \{r+\tfrac{\theta^2}{2}\,\partial_rV\}^{\frac12}\ A\ \{r+\tfrac{\theta^2}{2}\,\partial_rV\}^{-\frac12}=
        -\left[1+\frac{\theta^2}{2}\,\frac{\partial_rV}{r}\right]\partial^2_r+V_{{\rm eff}}\,,
\end{equation}
where
\begin{equation}\label{pveff}
        V_{{\rm eff}}=V+\frac{l^2}{r^2}\left[1+\frac{\theta^2}{2}\,\partial_r^2 V\right]
        +U\,.
\end{equation}
The term denoted by $U$ is a function of $r$, independent of $l$, which corresponds to $U=-1/4r^2$ for $\theta=0$. From these expressions one can read the dependence of the effective background on the angular momentum for a step-like function $V\left(r^2\right)$ or, eventually, a smooth approximation to it. Indeed, expression \eqref{pveff} shows that, to leading order in $\theta$, the centrifugal barrier term has an extra contribution proportional to $\theta^2\partial^2_rV$. For a step-like background --or a smooth approximation to it-- this term is represented inside the disc by a positive function highly peaked near the boundary. This contribution is responsible for the repulsion at the boundary experienced by the oscillations with non-vanishing angular momentum.

\section{Casimir energy of the NC disc}\label{casim}

In this Section we use the results of Section \ref{nc-well} to study the energy due to the vacuum oscillations of a scalar field defined on the NC disc. We are therefore interested in the oscillation modes $\psi_n(t,x)$, which are given by
\begin{equation}
    \psi_n(t,x)=e^{-i\omega_n t}\psi_n(x)\,,
\end{equation}
where $\omega_n$ is the oscillation frequency and $\psi_n(x)$ is a solution to (see eq.\ \eqref{olqf})
\begin{equation}
    \delta^2S\cdot e^{-i\omega_n t}\psi_n(x)=\left\{\omega_n^2-m^2-A^\infty_N\right\}\psi_n(x)=0\,.
\end{equation}
From the eigenvalues of the operator $A^\infty_N$, given by expression \eqref{ncspec}, we obtain the frequencies
\begin{equation}
    \omega^{\pm l}_k:=\sqrt{m^2+\frac{\lambda^l_k}{\theta}}\,,
\end{equation}
where $L^l_{N-l}(\lambda^l_k)=0$, for $l=0,1,\ldots,N-1$ and $k=1,2,\ldots,N-l$.

The Casimir energy $E_{NC}$ is given by the sum of the ground state energies of the oscillation modes
\begin{align}\label{casi}
    E_{NC}&=\frac12 \sum_{k=1}^{N}\omega^0_k+\sum_{l=1}^{N-1}\sum_{k=1}^{N-l}\omega^l_k\\\nonumber
    &=\frac{1}{2\sqrt{\theta}} \sum_{k=1}^{N}\sqrt{\lambda^0_k+\theta\, m^2}+
    \frac{1}{\sqrt{\theta}} \sum_{l=1}^{N-1}\sum_{k=1}^{N-l}\sqrt{\lambda^l_k+\theta\, m^2}\,.
\end{align}
Since the Hilbert space of quantum oscillations is finite dimensional the Casimir energy of the NC disc is finite; however, as expected, it diverges in the commutative limit $N\rightarrow \infty$.

In particular, for the massless case we obtain
\begin{align}\label{casim=0}
    E_{NC}&=\frac{1}{2\sqrt{\theta}} \sum_{k=1}^{N}\sqrt{\lambda^0_k}+
    \frac{1}{\sqrt{\theta}} \sum_{l=1}^{N-1}\sum_{k=1}^{N-l}\sqrt{\lambda^l_k}\\\nonumber
    &\simeq \frac{c}{R}\,N^3+\ldots\qquad ({\rm for\ large\ }N)\,.
\end{align}
The asymptotic limit $E_{NC}\sim N^3$ for large $N$ is derived from the following bounds on the zeros of the Laguerre polynomials \cite{Gatteschi,IS}:
\begin{equation}
    \frac{2k+l+1}{\sqrt{N-l/2+1/2}}>\sqrt{\lambda^l_k}>\frac{\pi k+l-1/2}{\sqrt{4N-2l+2}}\,.
\end{equation}
A rough estimation based on these bounds gives $0.69<c<1.11$.

On the other hand, the large mass limit of expression \eqref{casi} reads
\begin{align}
    E_{NC}&=
    \frac{m}{2} \sum_{k=1}^{N}\left\{1+\frac{\lambda^0_k}{2\theta m^2}+\ldots\right\}+
    m \sum_{l=1}^{N-1}\sum_{k=1}^{N-l}\left\{1+\frac{\lambda^l_k}{2\theta m^2}+\ldots\right\}\\\nonumber
    &=\frac{1}{2}\,m \,N^2+\frac{1}{4\theta m}N^3+\ldots
\end{align}
After subtracting the first term --which represents the rest energy of the $N^2$ oscillators-- one obtains, as expected, that the vacuum energy vanishes for an infinitely heavy scalar particle. Subsequent terms, depending on higher powers of $(\theta m^2)^{-1}$, can be obtained from the relations between the homogeneous sums of products of the zeros of a Laguerre polynomial and its coefficients.

Casimir energies calculations have been performed for $T^2_F$ \cite{Chaichian:2001pw} and $S^2_F$ \cite{Demetrian:2002pm}. In the case of the fuzzy torus --which is characterized by a rational noncommutativity parameter $\theta=N^{-1}$, with $N\in\mathbb{Z}^+$--, the orthonormal modes span an $N^2$-dimensional Hilbert space and its Casimir energy, which is finite, grows as $N^3/4\pi$ as $N$ increases. On the other hand, the normal modes of a scalar field on $S^2_F$ span a $J$-dimensional Hilbert space, with $J$ related to a particular representation of $su(2)$. As in our model, the Casimir energy is finite and diverges with increasing $J$.

\section{Conclusions}\label{s5}

We have studied a scalar field in 2+1 dimensions that interacts with a rotationally invariant background by means of a space-like Moyal product. In a certain limit one obtains the spectrum of oscillations of the scalar field confined in a NC disc. The definition of the Moyal product in a manifold with boundaries presents some obstacles --discovered in NC Chern-Simons theories \cite{Pinzul:2001qh,Lugo:2001ya}-- related to the appropriate choice of boundary conditions. Our definition of a scalar field in a NC disc as the limit of the solutions in interaction with a confining background circumvents these difficulties for in our setting the Moyal product corresponds to the algebra of functions on the whole NC plane.

We have shown that, under this procedure, the oscillations of the field in the NC disc span a finite-dimensional Hilbert space whose dimension is given by the relation between the area of the disc and the minimal area in the Moyal plane. The number of states also coincides with the number of fundamental cells contained in the finite physical volume in phase space. Moreover, the oscillation modes manifest the rotational and parity symmetries of the disc. We have also shown that, at some point, as the angular momentum of the modes increases the particle moves back towards the center of the disc. From a perturbative point of view, this can be interpreted as the effect of a repulsion at the boundary due to a term proportional to $\theta^2 l^2\,\partial^2_rV/r^2$ --where $V$ is the confining background-- arising from a small-$\theta$ expansion of the Moyal product.

Motivated by its application in the context of NC Kaluza-Klein scenarios we considered the Casimir energy $E_{NC}$ in the NC disc. Since the oscillations of the quantum field span a finite-dimensional Hilbert space, the Casimir energy is finite and needs no regularization. We determined the divergence of $E_{NC}$ as the noncommutativity parameter tends to zero and we obtained analogous results as for the fuzzy sphere and torus. However, in either of these NC spaces it is not clear how to get the commutative result from the $\theta\rightarrow 0$ limit. As mentioned in the Introduction, the finite Casimir energy arising from noncommutative KK oscillation modes can turn from a negative to a positive contribution to the dark energy \cite{Fabi:2006rs}. This is a consequence of the mechanism of absorption of infinities into the vacuum energy of Minkowski spacetime: on NC models some of these ``infinities'' are already ``regularized'' so that the effective contribution to the vacuum energy can change sign.

The procedure studied in this article provides an alternative point of view for the truncation used to define the fuzzy disc algebra \cite{Lizzi:2003ru}. After an appropriate identification of the corresponding parameters, the oscillation spectrum we have derived coincides with the eigenvalues of the Laplacian on the fuzzy disc. Nevertheless, we remark that since our formulation does not rely entirely on the symmetries of the disc (though the explicit resolution of the problem does) it opens the possibility of studying other regions with boundary on the Moyal plane. Work along these lines is currently in progress.

\acknowledgments{S.A.F.V.\ and F.V.\ acknowledge support from CONICET, Argentina. This work was partially supported by grants from CONICET (PIP 01787), ANPCyT (PICT-2011-0605) and UNLP (Proy.~11/X615), Argentina.}

\appendix

\section{Bound states for finite $\Lambda$}\label{thewell}

If we replace expansion \eqref{expansion} into equation \eqref{eigequ} we obtain the following recurrence relations for the coefficients $c^l_n(\lambda)$:
\begin{align}\label{rec}
    \sqrt{l+1}\,c^l_1(\lambda)&=\left(1+l-\lambda+\Lambda\cdot\Theta(l-N)\right)c^l_0(\lambda)\,,
    \\\nonumber
    \sqrt{(n+l)n}\,c^l_n(\lambda)&=-\sqrt{(n+l-1)(n-1)}\,c^l_{n-2}(\lambda)+\\\nonumber
    &\mbox{}+\left(2n-1+l-\lambda+\Lambda\cdot\left\{\Theta(n+l-1-N)+\Theta(n-1-N)\right\}\right)c^l_{n-1}(\lambda)\,,
\end{align}
for $n\geq 2$. The solution to these recurrence relations will be written in terms of the functions
\begin{equation}
    u^l_n(z):=\frac{\sqrt{n!\,l!}}{\sqrt{(n+l)!}}\, L^l_n(z)\,;\qquad
    v^l_n(z):=\frac{\sqrt{n!(n+l)!}}{\sqrt{l!}}\, U(n+1,1-l;-z)\,,
\end{equation}
where $l,n\in\mathbb{Z}^+$ and $U(a,b;z)$ represents the confluent hypergeometric function \cite{A-S}. Due to the step functions in eqs.\ \eqref{rec} it is convenient to consider separately the following four cases, according to the absolute value of the angular momentum: (\ref{1}) $l=0$, (\ref{2}) $l=1$, (\ref{3}) $2\leq l\leq N-1$, (\ref{4}) $l\geq N$.

\subsection{The case $l=0$}\label{1}
The solution to the recurrence relations \eqref{rec} for $l=0$ is given by (up to an overall normalization constant)
\begin{equation}\label{sol0}
    c^0_n(\lambda)=\left\{\begin{array}{ll}
    u^0_n(\lambda)  &
    {\rm for\ }n=0,1,\ldots,N\\ \\
    \frac{u^0_{N}(\lambda)}{v^0_{N}(\lambda-2\Lambda)}\, v^0_n(\lambda-2\Lambda)   &
    {\rm for\ }n\geq N+1\,.
    \end{array}\right.
\end{equation}
Moreover, recurrence relations \eqref{rec} are satisfied if the following condition holds
\begin{equation}\label{spe0}
    u^0_N(\lambda)v^0_{N-1}(\lambda-2\Lambda)-
    u^0_{N-1}(\lambda)v^0_{N}(\lambda-2\Lambda)
    =0\,,
\end{equation}
for $\lambda<2\Lambda$. Eq.\ \eqref{spe0} determines the eigenvalues $\lambda$ (in units of $\theta^{-1}$) of the operator $A^\Lambda_N$ on the subspace $\mathcal{F}^0$ of states with vanishing angular momentum.

\subsection{The case $l=1$}\label{2}
The solution to the recurrence relations \eqref{rec} for $l=1$ is given by (up to an overall normalization constant)
\begin{equation}\label{sol1}
    c^1_n(\lambda)=\left\{\begin{array}{ll}
    u^1_n(\lambda)  &
    {\rm for\ }n=0,1,\ldots,N-1
    \\ \\
    \frac{v^1_{N}(\lambda-2\Lambda)}{v^1_{N-1}(\lambda-2\Lambda)}\,u^1_{N-1}(\lambda) &
    {\rm for\ }n=N
    \\ \\
    \frac{u^1_{N-1}(\lambda)}{v^1_{N-1}(\lambda-2\Lambda)}\, v^1_n(\lambda-2\Lambda)    &
    {\rm for\ }n\geq N+1\,.
    \end{array}\right.
\end{equation}
Moreover, recurrence relations \eqref{rec} are satisfied if the following condition holds
\begin{equation}\label{spe1}
    \left[u^1_{N}(\lambda)+\frac{\Lambda}{\sqrt{N(N+1)}}\,u^1_{N-1}(\lambda)\right]v^1_{N-1}(\lambda-2\Lambda)-
    u^1_{N-1}(\lambda)v^1_{N}(\lambda-2\Lambda)
    =0\,,
\end{equation}
for $\lambda<2\Lambda$. Eq.\ \eqref{spe1} determines the eigenvalues $\lambda$ (in units of $\theta^{-1}$) of $A^\Lambda_N$ on $\mathcal{F}^{\pm 1}$.

\subsection{The case $2\leq l\leq N-1$}\label{3}
The solution to the recurrence relations \eqref{rec} for $2\leq l\leq N-1$ is given by (up to an overall normalization constant)
\begin{equation}\label{soll}
    c^l_n(\lambda)=\left\{\begin{array}{ll}
    u^l_n(\lambda)  &
    {\rm for\ }n=0,1,\ldots,N-l
    \\ \\
    A_l(\lambda)\, u^l_{n}(\lambda-\Lambda)-B_l(\lambda)\, v^l_n(\lambda-\Lambda)   &
    {\rm for\ }n=N-l+1,\ldots,N
    \\ \\
    C_l(\lambda)\, v^l_n(\lambda-2\Lambda) &
    {\rm for\ }n\geq N+1\,,
    \end{array}\right.
\end{equation}
where the coefficients $A_l(\lambda),B_l(\lambda),C_l(\lambda)$ are given by
\begin{align}
    \nonumber A_l(\lambda)&:=\sqrt{N(N-l)}
    \left[u^l_{N-l}(\lambda)v^l_{N-l-1}(\lambda-\Lambda)-u^l_{N-l-1}(\lambda)v^l_{N-l}(\lambda-\Lambda)\right]\,,
    \\
    \nonumber B_l(\lambda)&:=\sqrt{N(N-l)}
    \left[u^l_{N-l}(\lambda)u^l_{N-l-1}(\lambda-\Lambda)-u^l_{N-l-1}(\lambda)u^l_{N-l}(\lambda-\Lambda)\right]\,,
    \\
    C_l(\lambda)&:=\frac{u^l_N(\lambda-\Lambda)}{v^l_N(\lambda-2\Lambda)}A_l(\lambda)-
    \frac{v^l_N(\lambda-\Lambda)}{v^l_N(\lambda-2\Lambda)}B_l(\lambda)\,.
\end{align}
Moreover, recurrence relations \eqref{rec} are only satisfied if the following condition holds
\begin{align}\label{spel}
    \left[u^l_{N}(\lambda-\Lambda)v^l_{N-1}(\lambda-2\Lambda)-u^l_{N-1}(\lambda-\Lambda)v^l_{N}(\lambda-2\Lambda)\right]
    \, A_l(\lambda)=\\\nonumber
    \left[v^l_{N}(\lambda-\Lambda)v^l_{N-1}(\lambda-2\Lambda)-v^l_{N-1}(\lambda-\Lambda)v^l_{N}(\lambda-2\Lambda)\right]
    \, B_l(\lambda)\,,
\end{align}
for $\lambda<2\Lambda$. Eq.\ \eqref{spel} determines the eigenvalues $\lambda$ (in units of $\theta^{-1}$) of $A^\Lambda_N$ on the subspaces $\mathcal{F}^{\pm l}$ for $2\leq l\leq N-1$.

\subsection{The case $l\geq N$}\label{4}
Finally, the solution to the recurrence relations \eqref{rec} for $l\geq N$ is given by (up to an overall normalization constant)
\begin{equation}\label{solgeqN}
    c^l_n(\lambda)=\left\{\begin{array}{ll}
    u^l_n(\lambda-\Lambda)  &
    {\rm for\ }n=0,1,\ldots,N
    \\ \\
    \frac{u^l_N(\lambda-\Lambda)}{v^l_N(\lambda-2\Lambda)}\, v^l_n(\lambda-2\Lambda) &
    {\rm for\ }n\geq N+1\,.
    \end{array}\right.
\end{equation}
Moreover, recurrence relations \eqref{rec} are satisfied if the following condition holds
\begin{eqnarray}\label{spegeqN}
    u^l_{N}(\lambda-\Lambda)v^l_{N-1}(\lambda-2\Lambda)-u^l_{N-1}(\lambda-\Lambda)v^l_{N}(\lambda-2\Lambda)=0\,,
\end{eqnarray}
for $\lambda<2\Lambda$. Eq.\ \eqref{spegeqN} determines the eigenvalues $\lambda$ (in units of $\theta^{-1}$) of $A^\Lambda_N$ on the subspaces $\mathcal{F}^{\pm l}$ for $l\geq N$.

\bigskip

One can explicitly check that the limit $\Lambda\rightarrow\infty$ of the results in this Appendix confirm expressions \eqref{ncspec} and \eqref{eigenfunctions}.


\end{document}